\documentclass[proceedings,conferenceproceedings,submit,moreauthors,pdftex,10pt,a4paper]{mdpi} 

\usepackage{hyperref}
\firstpage{1} 
\pubvolume{xx}
\issuenum{1}
\articlenumber{5}
\pubyear{2018}
\copyrightyear{2018}
\history{Received: date; Accepted: date; Published: date}

\Title{Overview on heavy-flavour at RHIC and LHC$^{\dagger}$}

\Author{Andrea Dubla$^{1,}$$^{2}$$\orcidA{}$*}

\AuthorNames{Andrea Dubla}

\address{%
$^{1}$ \quad Physikalisches Institut,
  Universit{\"a}t Heidelberg, 69120 Heidelberg,
  Germany \\
$^{2}$ \quad GSI Helmholtzzentrum f{\"u}r
  Schwerionenforschung, 64248 Darmstadt, Germany }

\corres{Correspondence: andrea.dubla@cern.ch}

\firstnote{Presented at Hot Quarks 2018 - Workshop for young scientists on the physics of ultrarelativistic nucleus-nucleus collisions, Texel, The Netherlands, September 7-14 2018}

\abstract{
The PHENIX and STAR Collaborations at the Relativistic Heavy-Ion Collider and ALICE, CMS, ATLAS and LHCb Collaborations at the Large Hadron Collider have measured the production of charmonium and bottonium states as well as open heavy flavor hadrons via their hadronic and semi-leptonic decays at mid-rapidity and in the semi-muonic decay channel at forward rapidity in pp, p--A and A--A collisions in an energy domain that ranges from $\sqrt{s}$ = 0.2 TeV to $\sqrt{s}$ = 13 TeV in pp collisions and from $\sqrt{s_{\rm NN}}$ = 0.2 TeV to $\sqrt{s_{\rm NN}}$ = 5.02 TeV in A--A collisions. In this contribution the latest experimental results will be reviewed.}


\begin{document}

\section{Introduction}
The lattice QCD calculations have shown that, in case the temperature exceeds a critical value \cite{lqcdl}, the nuclear matter undergoes a phase transition from colourless hadronic matter (baryons and mesons) to the so-called Quark Gluon Plasma (QGP), a de-confined system of quarks and gluons.
Heavy-flavour hadrons are unique probes for the investigation of the QGP. Open charm and beauty hadrons are expected to be sensitive to the energy density through the in-medium energy loss of their heavy quark constituents. In particular, charm and beauty quarks, due to the large masses, are produced at the early stage of the collision \cite{earlystage} and expected to loose less energy than light quarks and gluons while traversing the QGP due to dead-cone and colour-charge effects \cite{deadcone, colourcharge}. 
In addition quarkonium states are expected to be sensitive to the initial temperature of the system via their dissociation due to color screening. At Large Hadron Collider (LHC) energy scale the recombination effects, that can be pictured as in- medium recombination and statistical recombination at the freeze-out, are expected to become competitive with the dissociation producing a visible effect on the nuclear modification factor once compared with the same observable at lower collision energy (i.e. Relativistic Heavy-Ion Collider (RHIC)).
In order to exploit the sensitivity of heavy-flavour observables to medium effects 
a precise reference, where such effects are not expected, is needed and it is provided by pp collisions. 
The modification of the $p_{\rm T}$-differential yield in heavy-ion collisions with 
respect to pp collisions at the same centre-of-mass energy is quantified by the nuclear 
modification factor $R_{\rm AA}$.
Further insight into the transport properties of the medium is provided by the measurement of the elliptic flow $v_{2}$ of heavy-flavour particle.
At low $p_{\rm T}$, the $v_{2}$ of heavy-flavour particle is sensitive to the degree
of thermalization of charm and beauty quarks in the de-confined medium and to different hadronisation mechanisms, namely the fragmentation in the vacuum and the
coalescence in the medium.  At higher $p_{\rm T}$, the measurement of $v_{2}$
carries  information  on  the  path-length dependence of in-medium parton energy loss.

\section{Open heavy-flavour production in nucleon--nucleon collisions}

\begin{figure}[htb]
\begin{minipage}[b]{0.4\linewidth}
\centering
\includegraphics[height=2.2in]{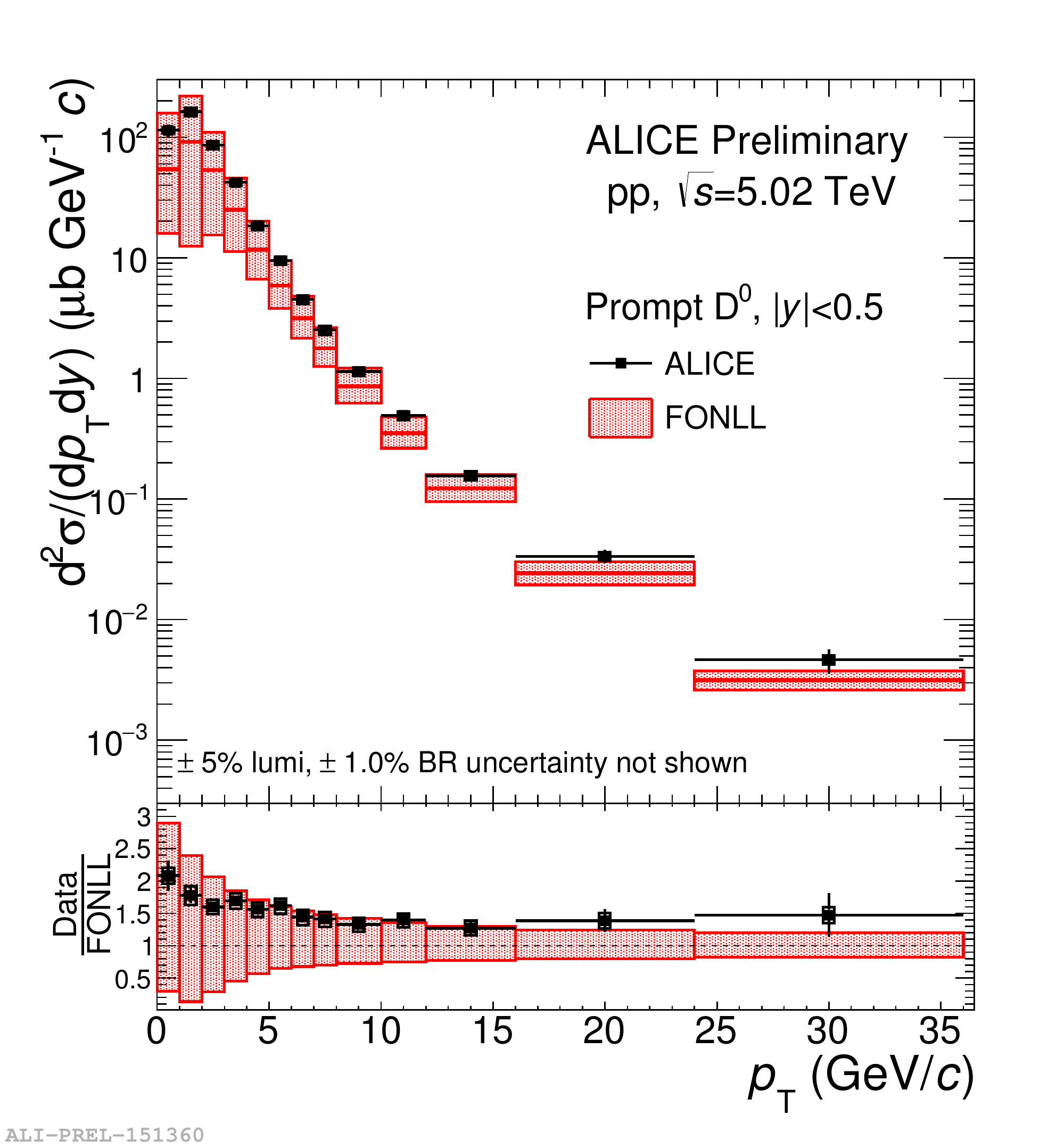}
\end{minipage}
\hspace{1cm}
\begin{minipage}[b]{0.5\linewidth}
\centering
\includegraphics[height=2.2in]{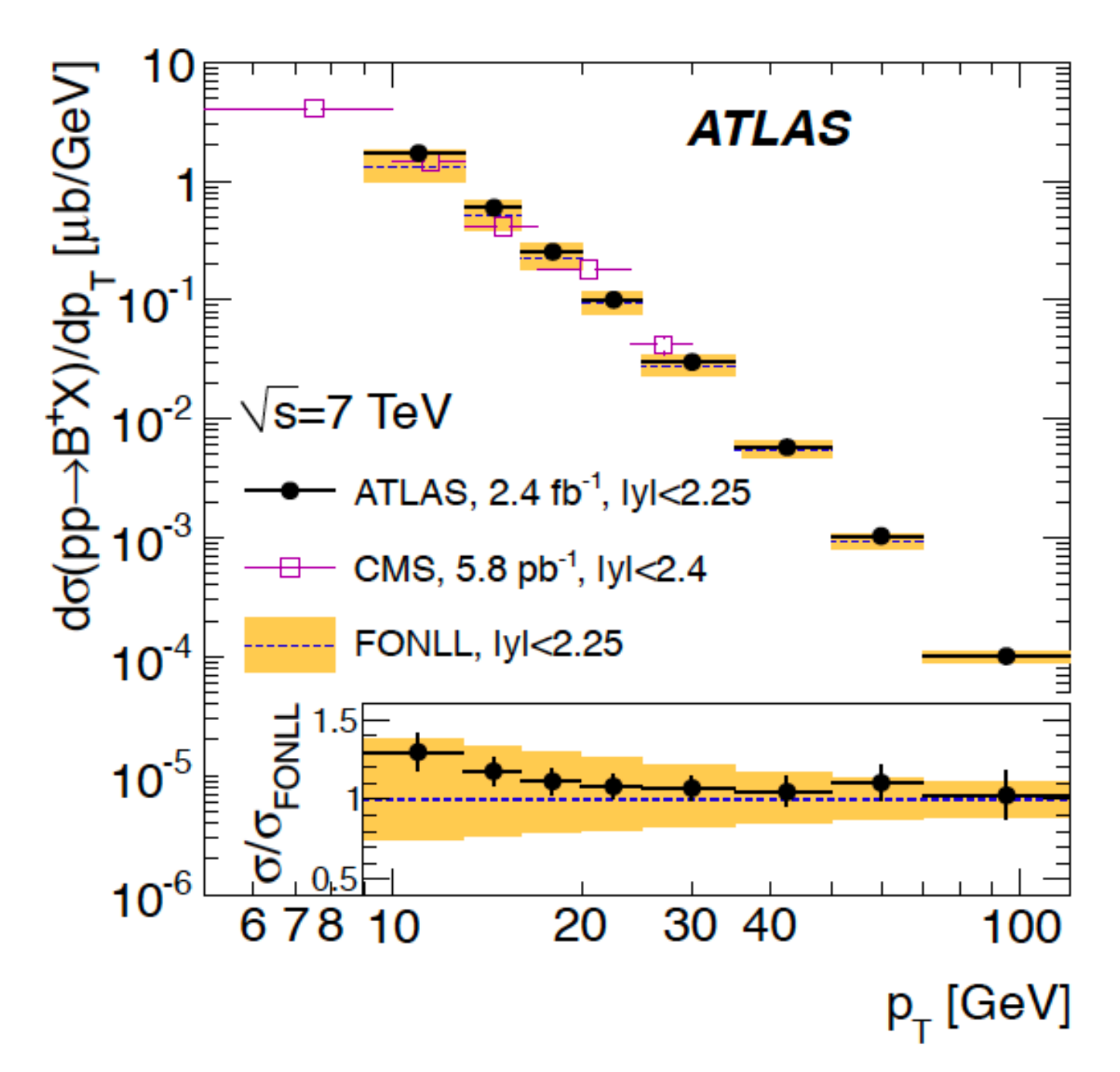}
\end{minipage}
\caption{ Left: $p_{\rm T}$-differential production cross sections for prompt $\rm D^{0}$ mesons in pp collisions at $\sqrt{s}$ = 5.02 TeV compared to FONLL calculations. Right: Differential cross-section of $\rm B^{+}$ production as a function of $p_{\rm T}$. The solid (open) circle points correspond to the measurement of ATLAS (CMS). Predictions of the FONLL calculation are compared to the experimental data.}
\label{fig1}
\end{figure}

At the RHIC and LHC the production of open heavy-flavour hadrons has been measured via their hadronic and semi-leptonic decays at mid- and forward-rapidity in pp collisions in an energy domain that rages from $\sqrt{s}$ = 0.2 TeV to $\sqrt{s}$ = 13 TeV.
In the left panel of Fig. \ref{fig1} presents the $\rm D^{0}$ meson measurements compared to FONLL \cite{FONLL} pQCD calculations in pp collisions at 5 TeV \cite{alicepp5tev}. The $\rm D^{0}$,  $\rm D^{+}$ and  $\rm D^{*+}$ d$\sigma$/d$p_{\rm T}$ are reproduced by the theoretical calculations within uncertainties \cite{alicepp7tev}. Yet, FONLL and POWHEG \cite{poweg} predictions tend to underestimate the data, whereas GM-VFNS \cite{GVFMS} calculations tend to overestimate the data.
The LHCb collaboration recently reported results on the $p_{\rm T}$-differential production cross section of D mesons at $\sqrt{s}$ = 13 TeV \cite{JHEP03} and found them to be in agreement with NLO predictions.
Studies of open-beauty production have also been performed in  hadronic \cite{cmsb, atlasb} and semi-leptonic decays \cite{aliceb, aliceb2} channels. As example, the right panel of Fig. \ref{fig1} presents the $\rm B^{+}$ $p_{\rm T}$ differential
cross section in pp collisions at  $\sqrt{s}$ = 7 TeV compared to FONLL predictions \cite{cmsb, atlasb}.
FONLL provides a good description of the experimental data. A similar agreement with FONLL is observed for the recent measurements of non-prompt
J/$\Psi$ cross section at $\sqrt{s}$ = 13 TeV \cite{JHEP1510}.

\section{Open heavy-flavour production in nucleus--nucleus collisions}

\begin{figure}[htb]
\begin{minipage}[b]{0.4\linewidth}
\centering
\includegraphics[height=2.2in]{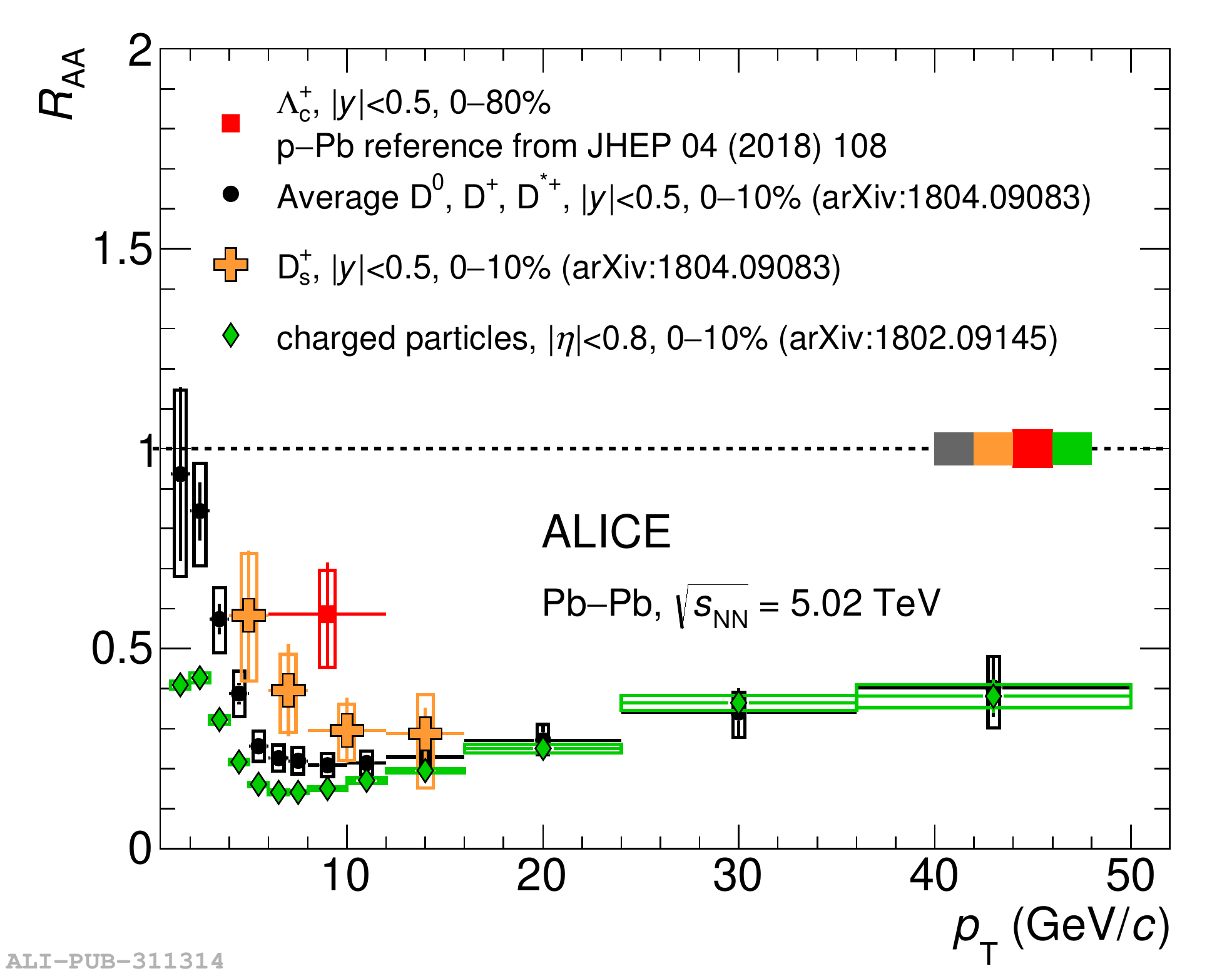}
\end{minipage}
\hspace{1cm}
\begin{minipage}[b]{0.5\linewidth}
\centering
\includegraphics[height=2.3in]{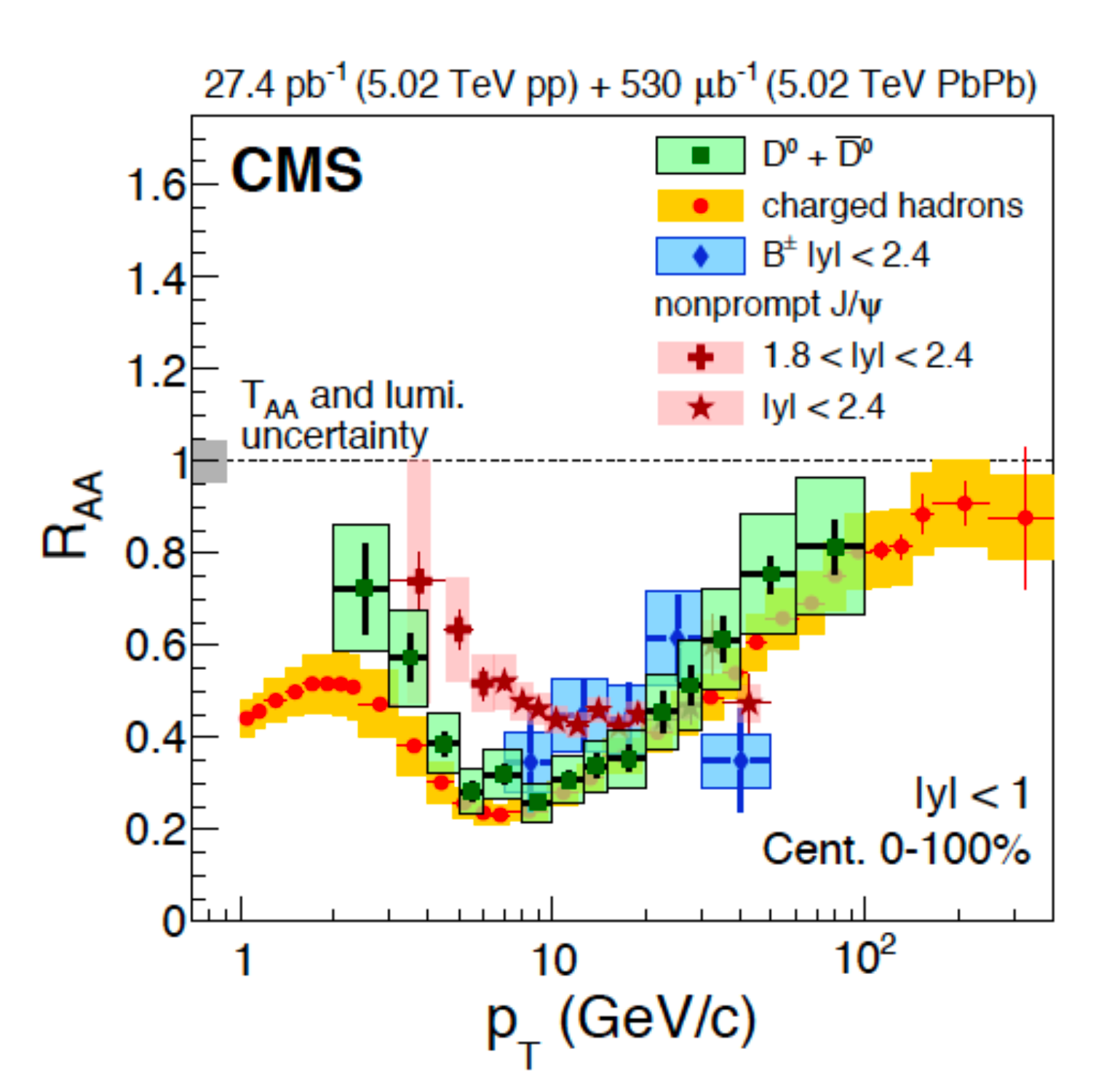}
\end{minipage}
\caption{ Left: : $R_{\rm AA}$ of prompt $\Lambda^{+}_{c}$ (red marker)
compared with the non-strange D mesons (black markers),
$\rm D^{+}_{s}$ (orange markers), and charged particle (green markers). Right: $\rm D^{0}$ meson $R_{\rm AA}$ as a function of $p_{\rm T}$ (green markers) compared to the $R_{\rm AA}$ of charged particles (red circles), $\rm B^{+}$ mesons
(blue markers), non-prompt J/$\Psi$ meson (purple markers) and charged particles (red markers).}
\label{fig2}
\end{figure}

Due to the QCD nature of parton energy loss discussed in the introduction, hierarchy in the  $R_{\rm AA}$ is expected to be observed when comparing the mostly gluon-originated light-flavor hadrons to D and to B mesons: $R_{\rm AA}$($\pi$) $<$ $R_{\rm AA}$(D) $<$ $R_{\rm AA}$(B). Clearly there are caveats to be taken into account while interpreting the results, like the different parton $p_{\rm T}$ spectra that can modify this picture.
The  left  panel of Fig. \ref{fig2} shows the $R_{\rm AA}$ of  prompt $\Lambda^{+}_{c}$ baryons measured in the 0--80\% centrality class (that is dominated by the 0--10\% production given the scaling of the yields with $N_{\rm coll} \cdot R_{\rm AA}$) compared with the average nuclear modification factors of non-strange D mesons, $\rm D^{+}_{s}$ mesons, and charged particles measured in the 0--10\% centrality class \cite{chpatalice, labdac, dmasonraaalice, cmsdo}. The $R_{\rm AA}$ of charged particles is smaller than that of D mesons by more than 2$\sigma$ of the combined statistical and systematic uncertainties up to $p_{\rm T}$ = 8 GeV/$c$, while they are compatible within 1$\sigma$ for $p_{\rm T}$ $>$10 GeV/$c$. The $R_{\rm AA}$ values of $\rm D^{+}_{s}$ mesons are larger than those of non-strange D mesons, but the two measurements are compatible within one standard deviation of the combined uncertainties.  A hint of a larger $\Lambda^{+}_{c}$ $R_{\rm AA}$ with respect to non-strange D mesons is observed, although the results are compared for different centrality classes.
The STAR and ALICE Collaborations also have studied the $\Lambda^{+}_{c}$/$\rm D^{0}$ ratio, which is essential to understand charm hadronisation \cite{labdacstar, labdac}.
The experimental results are described by the model calculation including coalescence \cite{coal1, coal2}. The curve
obtained by modelling charm hadronisation via vacuum fragmentation plus coalescence, which describes
the $\Lambda^{+}_{c}$/$\rm D^{0}$ratio measured in Au--Au collisions at RHIC energy, significantly underestimates the measurement in Pb--Pb collisions at the LHC.
The CMS $\rm D^{0}$  $R_{\rm AA}$ measured in the centrality range 0--100\% is compared in the right panel of Fig. \ref{fig2} to the CMS measurements of the $R_{\rm AA}$ of charged particles \cite{cmchpart}, $\rm B^{+}$ mesons \cite{cmsB} and non-prompt J/$\Psi$ meson \cite{cmsjpdsi} performed at the same energy and in the same centrality range.  The $\rm D^{0}$  $R_{\rm AA}$ was found to be
compatible with the $\rm B^{+}$  $R_{\rm AA}$ in the intermediate $p_{\rm T}$ region. The
$R_{\rm AA}$ of non-prompt J/$\Psi$, which was found to have almost no rapidity dependence, is shown
here measured in the $p_{\rm T}$ ranges 6.5--50 GeV/$c$ in $|y|$ $<$ 2.4, and 3--6.5 GeV/$c$ in 1.8 $<$ $|y|$ $<$ 2.4.
Its $R_{\rm AA}$ is found to be higher than the $\rm D^{0}$ meson $R_{\rm AA}$ in almost the entire $p_{\rm T}$ range. 

\section{Quarkonium measurements in nucleus-nucleus collisions}

\begin{figure}[htb]
\begin{minipage}[b]{0.4\linewidth}
\centering
\includegraphics[height=1.8in]{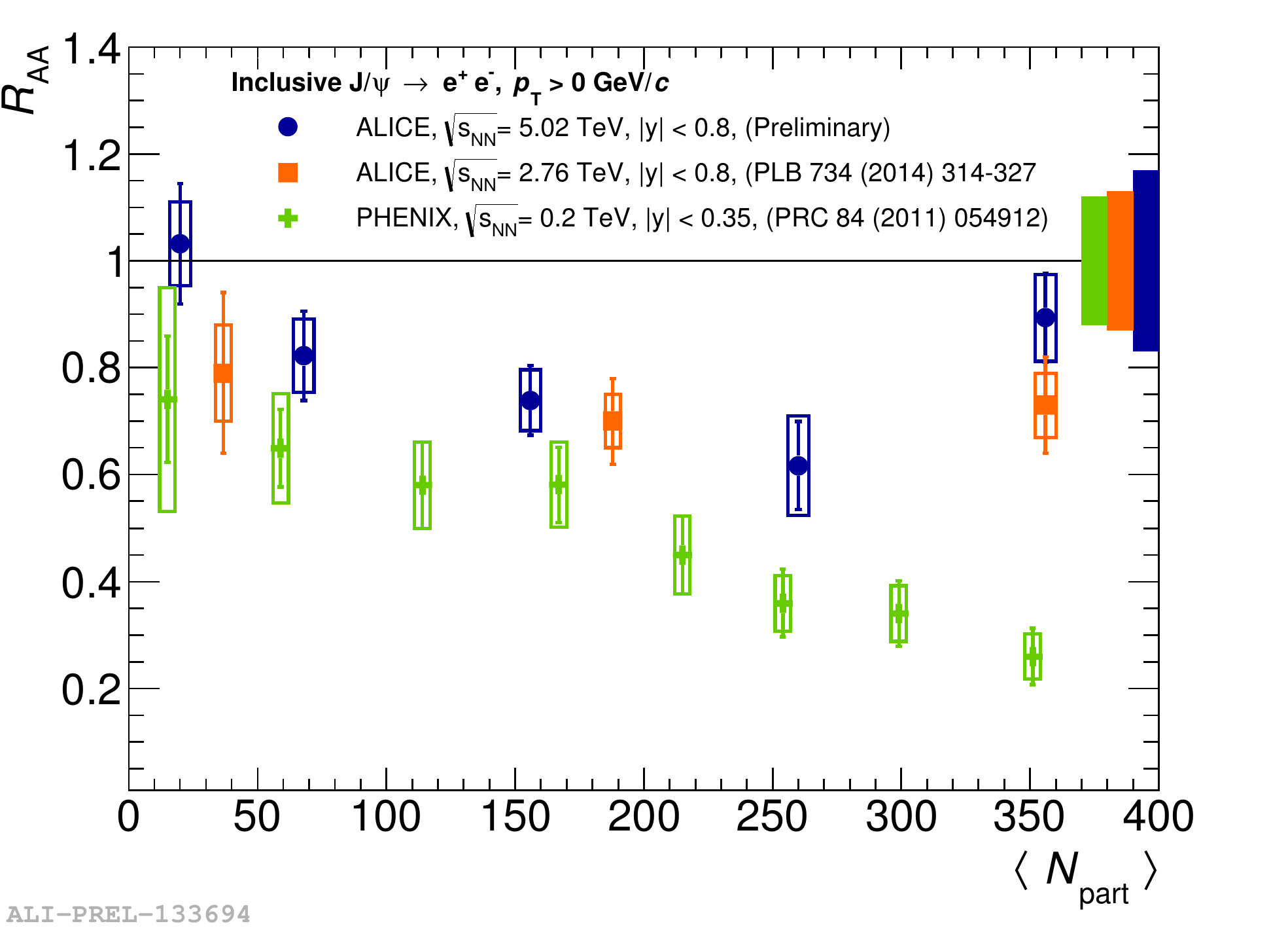}
\end{minipage}
\hspace{1cm}
\begin{minipage}[b]{0.5\linewidth}
\centering
\includegraphics[height=1.8in]{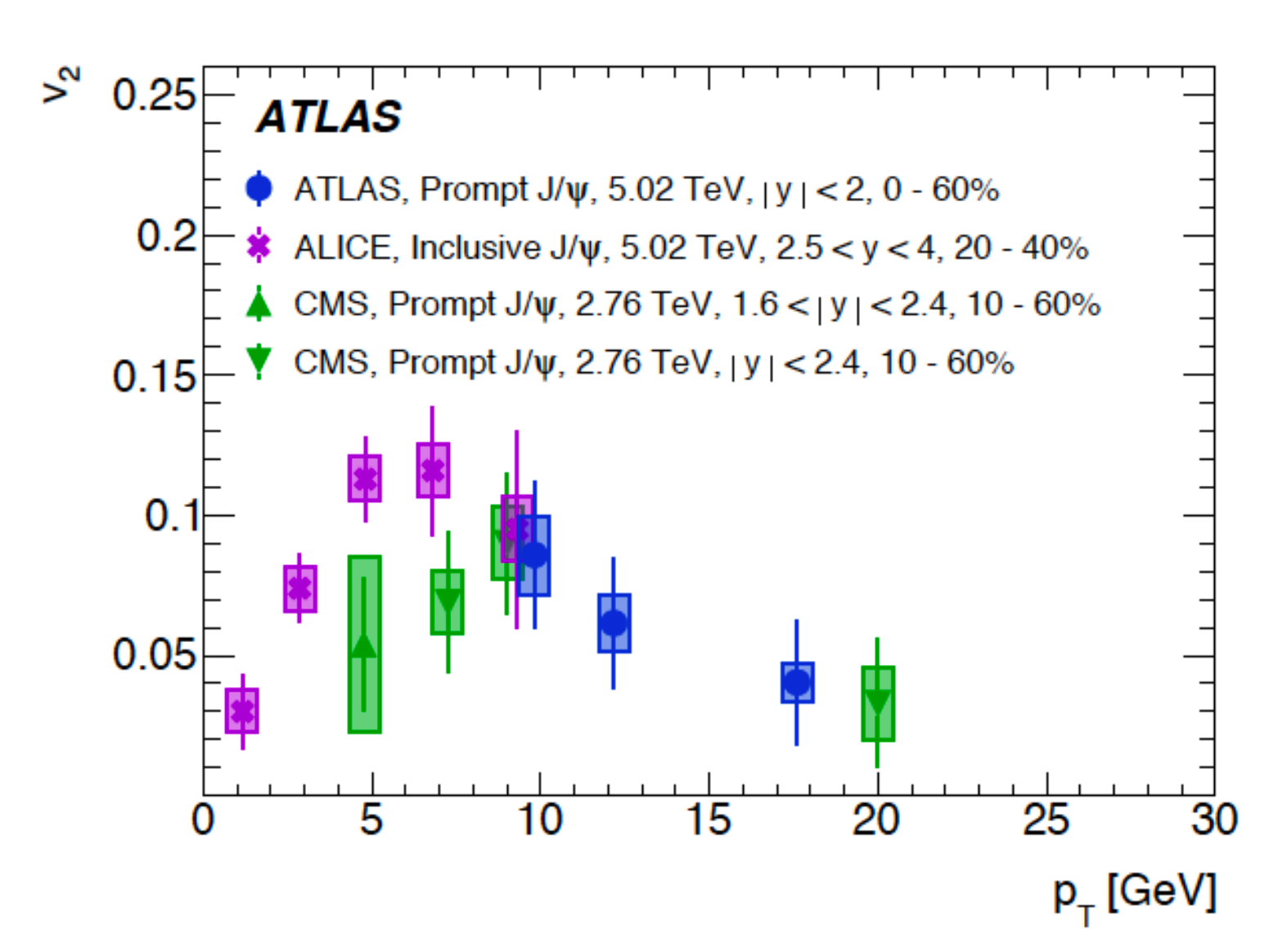}
\end{minipage}
\caption{ Left: Inclusive J/$\Psi$ $R_{\rm AA}$ at mid-rapidity as a function of centrality as measured by PHENIX and ALICE. Right: Results for $v_{2}$ as a function of $p_{\rm T}$ of prompt J/$\Psi$ as measured by ATLAS and CMS compared with inclusive J/$\Psi$ with as measured by ALICE.}
\label{fig3}
\end{figure}

Quarkonium is considered one of the main probes of the investigation of QGP. Its production in A--A collisions is expected to be significantly suppressed with respect to the pp yield, scaled by the number of binary nucleon-nucleon collisions. The origin of such a suppression is thought to be the colour screening \cite{Matsui}.
However, the simple picture of color screening needs to be complemented by recombination effects at LHC energies.
On this regard, the centrality dependence of the inclusive J/$\Psi$ production at low-$p_{\rm T}$ in A--A collisions has been measured at both RHIC and LHC  \cite{phnjpsi, alicejpsi, cmsjpdsi, atlasjpsi}. In the left panel of Fig. \ref{fig3} the inclusive J/$\Psi$ nuclear modification factor, as a function of the number of participant nucleons ($N_{part}$), is compared to the one measured by the PHENIX in a similar kinematic range. It is evident a smaller J/$\Psi$ suppression at LHC energies, with respect to RHIC. The smaller suppression at the LHC is assumed to be due to the larger $\rm c\bar{c}$ pair multiplicity which allows a larger recombination, resulting in a compensation of the suppression from colour screening. The measured behaviour is expected by the statistical model \cite{Andronic}, where the J/$\Psi$ yield is completely determined by the chemical freeze-out conditions and by the abundance of $\rm c\bar{c}$ pairs.
In the right panel of Fig. \ref{fig3} the elliptic flow results for inclusive J/$\Psi$ from the ALICE experiment \cite{alicejpsiflow} and
prompt J/$\Psi$ from the CMS experiment \cite{CMSjpsiflow} as well as the results from the ATLAS experiment \cite{atlasjpsiflow} are shown as a function of $p_{\rm T}$.
Despite different rapidity selections, the magnitudes of the elliptic flow coefficients are compatible with
each other. Two features can be observed: first, the hydrodynamic peak is around 7 GeV/$c$, a value that
is significantly higher than what is observed for charged particles where the peak is around 3 to
4 GeV/$c$. This effect can be described qualitatively by thermalisation of charm quarks in the quark gluon
plasma medium with J/$\Psi$ regeneration playing a dominant role in the flow formation. The second
feature is that $v_{2}$ has a substantial magnitude at high $p_{\rm T}$. This can be connected with the suppression
of J/$\Psi$  production due to mechanisms involving interactions with the medium such as energy loss \cite{atlasjpsiflow}.

\reftitle{References}



\end{document}